\documentclass[useAMS,usenatbib,usegraphicx]{mn2e}

\title[An environmental Butcher-Oemler effect in intermediate redshift X-ray clusters]{An environmental Butcher-Oemler effect in intermediate redshift X-ray clusters\thanks{Based on observations obtained with MegaPrime/MegaCam, a joint project of CFHT and CEA/DAPNIA, at the Canada-France-Hawaii Telescope (CFHT) which is operated by the National Research Council (NRC) of Canada, the Institut National des Science de l'Univers of the Centre National de la Recherche Scientifique (CNRS) of France, and the University of Hawaii. This work is based in part on data products produced at TERAPIX and the Canadian Astronomy Data Centre as part of the Canada-France-Hawaii Telescope Legacy Survey, a collaborative project of NRC and CNRS.}}
\author[S.A. Urquhart, J.P. Willis, H. Hoekstra, M. Pierre]{S.A. Urquhart$^{1}$\thanks{E-mail: surquhar@uvic.ca}, J. P. Willis$^{1}$, H. Hoekstra$^{1,2,3}$, M. Pierre$^{4}$\\$^{1}$Department of Physics and Astronomy, University of Victoria, Elliot Building, 3800 Finnerty Road, Victoria, BC, V8P 1A1, Canada.\\$^{2}$Alfred P. Sloan Fellow.\\$^3$Leiden Observatory, Leiden University, PO Box 9513, 2300 RA, Leiden, The Netherlands.\\$^4$Service d'Astrophysique, Bât. 709, CEA Saclay, 91191 Gif sur Yvette Cedex, France}

\begin{document}

\date{Accepted. Received; in original form}

\pagerange{\pageref{firstpage}--\pageref{lastpage}} \pubyear{2006}

\maketitle

\label{firstpage}

\begin{abstract}
  We present uniform CFHT Megacam $g$ and $r$ photometry for 34 X-ray
  selected galaxy clusters drawn from the X-ray Multi-Mirror (XMM)
  Large Scale Structure (LSS) survey and the Canadian Cluster
  Comparison Project (CCCP). The clusters possess well determined
  X-ray temperatures spanning the range $1<kT(\rm keV)<12$.  In
  addition, the clusters occupy a relatively narrow redshift interval
  ($0.15<z<0.41$) in order to minimize any redshift dependent
  photometric effects. We investigate the colour bimodality of the
  cluster galaxy populations and compute blue fractions using criteria
  derived from Butcher and Oemler (1984). We identify a trend to
  observe increasing blue fraction versus redshift in common with
  numerous previous studies of cluster galaxy populations.  However,
  in addition we identify an environmental dependence of cluster blue
  fraction in that cool (low mass) clusters display higher blue
  fractions than hotter (higher mass) clusters.  Finally, we
  tentatively identify a small excess population of extremely blue
  galaxies in the coolest X-ray clusters (essentially massive groups)
  and note that these may be the signature of actively star bursting
  galaxies driven by galaxy-galaxy interactions in the group
  environment.
\end{abstract}
\nokeywords

\section{Galaxy populations in clusters}

The observed properties of galaxy populations reflect the environment
in which they are located.  Comparisons of galaxy populations drawn
from low (the field) and high density (rich galaxy clusters)
environments indicate that the population distribution described using
measures such as current star formation rate (e.g. \citealt{balogh99},
\citealt{pogg06}), integrated colour (e.g. \citealt{blanton05}) and
morphology (e.g. \citealt{dressler97}, \citealt{treu03}) varies as a
function of changing environment. From studies such as these it is
clear that galaxies located in the cores of rich clusters display
lower star formation rates, redder colours and more bulge dominated
morphologies compared to galaxies located in the field.

Studies of the fraction of blue galaxies contributing to a galaxy
cluster provided some of the first direct evidence for the physical
transformation of galaxies in cluster environments. \cite{bo84}
reported an increase in the fraction of blue galaxies in 33 rich
galaxy clusters out to $z\sim0.5$ compared to local clusters. However,
subsequent studies designed to expand upon this initial discovery
highlighted the many complexities associated with this relatively
straightforward technique, including varying intrinsic cluster
properties with redshift (e.g. X-ray luminosity as discussed by
\citealt{andreon99}), the use of $k$-corrections to determine rest
frame colour distributions \citep{andreon05} and the challenge of
obtaining large samples with uniform photometry (successfully overcome
by \citealt{loh08}).

In addition to the Butcher-Oemler (BO) effect measured employing
optical galaxy colours, analogous BO-type effects have been reported
as either a morphological BO effect (increasing spiral fraction in
clusters with increasing redshift; \citealt{pogg99}) and an infra-red (IR) BO
effect (increasing fraction of dust enshrouded star forming falaxies
in clusters with increasing redshift; \citealt{duc02};
\citealt{saintonge08}) to name two examples.

An alternative approach is to consider an environmental Butcher-Oemler
effect whereby one attempts to determine the variation of blue
fraction as a function of varying intrinsic cluster properties
selected over narrow redshift intervals.  This has been achieved by
comparing blue fractions within clusters at increasing clustercentric
radii (e.g. \citealt{ellingson01}; \citealt{loh08}) or by considering
blue fractions measured between clusters of differing X-ray
luminosities (e.g. \citealt{wake05}).  A number of studies report the
decrease of the fraction of blue galaxies with decreasing scaled
clustercentric radius, e.g. the virial radius determined employing
either cluster dynamics \citep{ellingson01}, correlation properties
\citep{loh08}, or extrapolated from X-ray properties \citep{wake05}.

The currently favoured explanation for these observed trends is that
infalling field galaxies are processed physically as they travel from
the field, through the cluster outskirts and virialise in the central
cluster region \citep{berrier09}.  However, the extent to which
cluster galaxies were ``pre-processed'' by physical effects occuring
during an earlier residence in a galaxy group\footnote{Note that we
generally refer both groups and clusters of galaxies as "clusters" in
this paper and quantify our description using the X-ray temperature.}
remains contested \citep{li09}.  Numerous physical processes have been
suggested as the agents of this apparent transformation of galaxy
populations.  However, the dominant physical process(es) to which an
infalling galaxy is subject remains unclear.

The two principal examples of such processes are ram pressure
stripping and galaxy galaxy interactions.  Ram pressure stripping
describes the effective force experienced by the diffuse gas component
of the infalling galaxy as it travels through the hot, dense
intra-cluster medium (ICM; \citealt{gunn72}).  Both hot and cold gas
may be stripped from the infalling galaxy leading to the exhaustion of
the available gas supply that will otherwise cool and form stars
(\citealt{abadi99}; \citealt{kawata08}; \citealt{mccarthy08}).
Hydrodynamical simulations have indicated that the ICM associated with
galaxy group and cluster environments will strip of order 70\% to
100\% respectively of hot gas from a typical infalling spiral galaxy
within one crossing time (\citealt{kawata08};
\citealt{mccarthy08}). The effect is manifest as a sharp decline in
the galaxy star formation rate effective on a timescale comparable to
the rate at which the unstripped cold gas supply is consumed.  The
observations of a population of red (i.e. passive) spiral galaxies
contributing to the red sequence would appear to support the view that
some fraction of galaxies experience the stripping of disc gas via ram
pressure effects \citep{wolf09}.  The observation of galaxies
displaying extended H{\sevensize I} tails in the Virgo cluster
\citep{chung07} is nominally consistent with the expectations of ram
pressure stripping. However, we note that the authors also report the
presence of close companions to these galaxies in a number of cases
and comment that a combination of ram pressure and tidal stripping
provides a more compelling explanation.

When referring to galaxy-galaxy interactions we note that this may
indicate one of a wide range of encounters.  Interactions may be
predominantly tidal between close neighbours resulting in halo gas
being moved outward where it is more readily stripped
(e.g. \citealt{chung07}).  High speed encounters (either referred to
as harrassment or threshing in the literature) may also result in
tidal stripping of halo gas \citep{moore96}.  Finally, infalling
galaxies may merge with existing cluster members. The products of such
merger encounters may be predicted by considering the mass ratio of
the merging galaxies: large mass ratios result in enhanced star
formation in the satellite galaxy yet may not lead to a star burst in
the more massive companion \citep{cox08}.  Equal mass mergers on the
other hand result in the complete disruption of the infalling spiral
galaxy (for example) to form a bulge dominated system accompanied by a
significant central star burst \citep{matteo07}.  The internal
disruption associated with such strong interactions can result in a
short term enhancement in star formation followed by a rapid
exhaustion of the available cold gas supply.  The potential effect of
such strong galaxy-galaxy interactions is of interest to studies of
the environmental dependence of galaxy evolution as the merging rate
(the product of relative velocity and interaction cross section) is
predicted to be a strong function of environment: the cross section
for disruptive encounters increases as the cluster velocity dispersion
approaches the internal velocity of the infalling galaxy
\citep{makino97}.  Compelling evidence for enhanced galaxy-galaxy
interactions in rich cluster environments is provided by Hubble Space
Telescope (HST) observations indicating a high merger fraction in such
environments compared to field comparison samples
(\citealt{dressler94}; \citealt{vandokkum99}).  However, such
observations must be contrasted with mid-infrared selected moderate
starburst galaxies located in rich cluster environments whose optical
morphologies resemble undisturbed spiral galaxies (\citealt{geach09};
\citealt{oemler09}).

A picture is therefore emerging whereby multiple mechanisms (ICM
stripping, merger induced star formation and tidally induced star
formation) may participate in the physical processing of infalling
galaxies.  Currently unanswered questions focus upon whether more than
one physical process acts upon a typical galaxy falling into a dense
environment (and which might be considered dominant) and whether the
relative importance of each of the suggested physical processes
changes as a function of the global properties of the group/cluster
environment into which the galaxy falls.  In this paper we attempt to
answer the related question of whether the effectiveness of galaxy
processing can be determined as a function of global cluster
environment. Our approach is to compute the blue fraction as defined by
\cite{bo84} for a sample of X-ray selected galaxy clusters and to
determine whether the blue fraction displays a significant trend
versus X-ray temperature (here employed as a proxy for the global
cluster mass).

Throughout this paper, values of $\Omega_{\rm M,0}=0.3$,
$\Omega_{\Lambda,0}=0.7$ and $H_0=70 \rm \, kms^{-1} \, Mpc^{-1}$ are
adopted for the present epoch cosmological parameters describing the
evolution of a model Friedmann-Robertson-Walker universe. All
magnitude information is quoted using AB zero point values.

\section{The galaxy cluster sample}

\begin{figure}
\includegraphics[width=84mm]{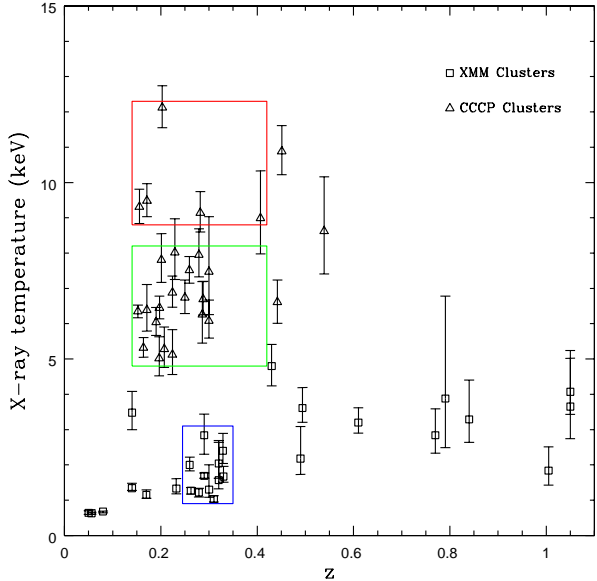}
\caption{Redshift and temperature distribution of the hot, mid and cool clusters samples as defined in the text. Open triangles represent CCCP clusters and open squares represent XMM-LSS clusters.}
\label{fig_zt}
\end{figure}

The data presented in this paper are drawn from two complementary
samples of X-ray selected galaxy clusters.  Clusters with X-ray
temperatures $\rm T<3$~keV are drawn from the X-ray Multi-Mirror (XMM)
Large Scale Structure (LSS) survey.  Clusters are selected from the 5
deg$^2$ ``Class 1'' sample of \cite{pacaud07} according to X-ray
temperature $\rm T<3$~keV and spectroscopic redshift $0.25<z<0.35$
(Figure \ref{fig_zt}).  Such relatively cool X-ray clusters are often
referred to as galaxy groups in the literature. However, all X-ray
selected systems considered in this paper are referred to as
``clusters'' for simplicity. The above criteria generate a sample of
11 clusters which are referred to as ``cool'' in the following
analysis.

Clusters with X-ray temperatures $\rm T>5$~keV are selected from the
sample of Horner (2001) and form part of the Canadian Cluster
Comparison Project (CCCP; \citealt{bildfell08}).  Although the final
CCCP sample will contain some 50 X-ray selected galaxy clusters, we
consider here the 27 systems with accompanying optical data obtained
using the Canada France Hawaii Telescope (CFHT) MegaCam imager (see
below). Due to the large temperature range covered by the CCCP sample
($5< \rm T(keV)<12$) we further subdivide the CCCP sample into ``mid''
clusters displaying $5<\rm T(keV)<8$ and ``hot'' displaying $\rm
T>8$~keV. We limit the CCCP sample to redshifts $0.15<z<0.4$ in order
to maximise the available sample size while reducing the redshift
interval over which photometric quantities are $k$-corrected to a
common epoch.  These considerations limit the size of the mid and hot
samples to 18 and 5 clusters respectively. The X-ray properties of all
clusters are shown in Table \ref{tab_sample}.

The CCCP clusters all lie above the nominal X-ray flux limit of the
XMM-LSS survey and cover the full range of scatter observed in scaling
relations such as the X-ray $L-T$ relation.  In this sense, the CCCP
clusters would all be detected by X-ray images of the same quality as
the XMM-LSS survey images (irrespective of how the CCCP cluster were
originally selected).  However, XMM-LSS lacks the areal coverage to
detect such massive clusters which display low sky surface densities.
Therefore, the combined XMM-LSS/CCCP sample is a representative,
(largely) unbiased sample of X-ray clusters with temperatures $1<kT
(\rm keV)<12$. We note that the XMM LSS clusters populating the cool
sample are moderately biased toward higher X-ray luminosities than the
average cluster population at that temperature (see \citealt{pacaud07}
Figure 8). This implies a moderate bias toward gas rich or centrally
condensed systems which we will recall at relevent points in the
following discussion.

The characteristic cluster scaling radii employed in this work are
based upon $r_{500}$.  This is defined as the radius at which the
enclosed cluster mass density equals 500 times the critical density of
the universe at the cluster redshift \citep{pacaud07}. Converting the
relation of \cite{fin01} to a $\Lambda$CDM cosmology, \cite{willis05}
define
\begin{equation}
{
r_{500}=0.375 ~{\rm T}^{0.63}h_{73}(z)^{-1} \rm Mpc
}
\end{equation}
where T measured X-ray temperature in keV and $h_{73}$ is the Hubble
constant in units of $73~ \rm kms^{-1}Mpc^{-1}$.

\begin{table}
\begin{centering}
 \caption{Properties of the cluster sample. Clusters are sorted with
   increasing temperature. Clusters possessing $T(\rm keV)<3$ are
   labelled ``Cool'', clusters possessing $5<T(\rm keV)<8$ are
   labelled ``Mid'' and those possessing $T(\rm keV)>8$ are labelled
   ``Hot''.}
\label{tab_sample}
\begin{tabular}{lccccc}
\hline
Cluster &R.A.&Dec.& $T_{X}$ & z & $r_{500}$ \\
& (deg.) & (deg.) & (keV) & & (kpc) \\
\hline

XLSSC 13 & 36.858 & -4.538 & $1.03^{+0.1}_{-0.08}$& 0.31 & 340 \\
XLSSC 51 & 36.498 & -2.826 & $1.22^{+0.12}_{-0.13}$& 0.28 & 384 \\
XLSSC 44 & 36.141 & -4.234 & $1.27^{+0.09}_{-0.1}$ & 0.26 & 399 \\
XLSSC 08 & 36.337 & -3.801 & $1.30^{+0.7}_{-0.22}$& 0.30 & 387 \\
XLSSC 40 & 35.523 & -4.546 & $1.57^{+1.07}_{-0.25}$ & 0.32 & 402 \\
XLSSC 23 & 35.189 & -3.433 & $1.67^{+0.29}_{-0.16}$& 0.33 & 497 \\
XLSSC 22 & 36.916 & -4.857 & $1.69^{+0.08}_{-0.07}$ & 0.29 & 472  \\
XLSSC 25 & 36.353 & -4.679 & $2.00^{+0.22}_{-0.17}$& 0.26 & 533 \\
XLSSC 18 & 36.008 & -5.090 & $2.04^{+0.65}_{-0.42}$& 0.32 & 615 \\
XLSSC 27 & 37.014 & -4.851 & $2.84^{+0.6}_{-0.54}$& 0.29 & 653 \\
XLSSC 10 & 36.843 & -3.362 & $2.40^{+0.49}_{-0.36}$ & 0.34 & 574 \\ \hline

MS0440+02 & 70.805 & 2.166 & $5.02^{+0.61}_{-0.5}$ & 0.19 & 957 \\
A1942 & 219.600 & 3.669 & $5.12^{+0.71}_{-0.56}$ & 0.22 & 957 \\
A0223 & 24.477 & -12.815 & $5.28^{+0.63}_{-0.52}$ & 0.21 & 964 \\
A2259 & 260.033 & 27.668 & $5.32^{+0.29}_{-0.27}$ & 0.16 & 1007 \\
A1246 & 170.9972& 21.482 & $6.04^{+0.42}_{-0.37}$ & 0.19 & 1078 \\
A2537 & 347.092 & -2.187 & $6.08^{+0.59}_{-0.49}$ & 0.30 & 1039 \\
A0959 & 154.433 & 59.556 & $6.26^{+0.93}_{-0.81}$ & 0.29 & 1022 \\
A0586 & 113.072 & 31.637 & $6.39^{+0.72}_{-0.6}$ & 0.17 & 1127 \\
A0115 & 13.980 & 26.422 & $6.45^{+0.33}_{-0.31}$ & 0.20 & 1120 \\
A0611 & 120.228 & 36.065 & $6.69^{+0.51}_{-0.44}$ & 0.29 & 1100 \\
A0521 & 73.510 & -10.244 & $6.74^{+0.5}_{-0.45}$ & 0.25 & 1106 \\
A2261 & 260.609 & 32.139 & $6.88^{+0.47}_{-0.41}$ & 0.22 & 1153 \\
A2204 & 248.192 & 5.574 & $6.97^{+0.18}_{-0.18}$ & 0.15 & 1247 \\
MS1008-12  & 152.632 & -12.652 & $7.47^{+1.56}_{-1.21}$ & 0.30 & 1169 \\
CL1938+54 & 294.555 & 54.159 & $7.52^{+0.38}_{-0.37}$ & 0.26 & 1200 \\
A0520 & 73.554 & 2.924 & $7.81^{+0.74}_{-0.64}$ & 0.20 & 1262 \\
A1758 & 203.205 & 50.538 & $7.95^{+0.74}_{-0.62}$ & 0.28 & 1186 \\
A2111 & 234.914 & 34.429 & $8.02^{+0.95}_{-0.77}$ & 0.23 & 1267 \\ \hline

A0851 & 145.746 & 46.9945 & $8.99^{+1.34}_{-1.01}$ & 0.41 & 1291 \\
A0697 & 130.740 & 36.3662 & $9.14^{+0.6}_{-0.54}$ & 0.28 & 1343 \\
A2104 & 235.029 & -3.3017 & $9.31^{+0.5}_{-0.47}$ & 0.16 & 1438 \\
A1914 & 216.512 & 37.8244 & $9.48^{+0.49}_{-0.45}$ & 0.17 & 1444 \\
A2163 & 243.920 & -6.1438 & $12.12^{+0.62}_{-0.57}$ & 0.20 & 1663 \\ \hline

\hline
\end{tabular}
\end{centering}
\end{table}

\subsection{Optical photometry}

\begin{figure}
\includegraphics[width=84mm]{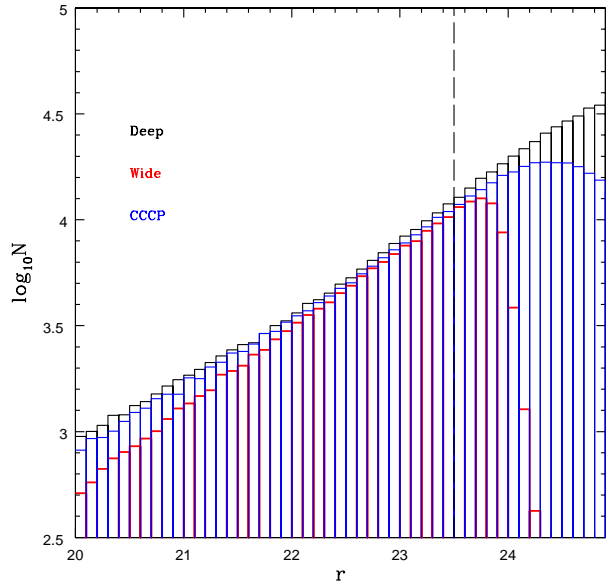}
\caption{Number counts as a function of 3\arcsec diameter $r$-band
  magnitudes in representative CCCP and CFHTLS Wide Megacam fields are
  compared to number counts in representative CFHTLS Deep Megacam
  fields. The vertical dashed line indicates $r=23.5$ and is the
  faintest magnitude employed in the following analysis.}
\label{fig_comp}
\end{figure}

Optical photometry for the XMM-LSS and CCCP cluster samples is
computed from CFHT Megacam images available for all fields.  XMM-LSS
clusters lie within the CFHT Legacy Survey (CFHTLS) wide synoptic
survey\footnote{http://www.cfht.hawaii.edu/Science/CFHTLS/} W1
area. The survey data consists of images taken in the CFHT $ugriz$
filter set.  Approximately 4 deg$^2$ of the XMM-LSS footprint lies
beyond the northern declination limit of the CFHTLS W1, of which 3
deg$^2$ has been imaged as part of the XMM-LSS follow-up campaign
using CFHT Megacam in the $grz$ bands. Image exposure times in these
additional fields are matched to the CFHTLS wide exposure times for
each filter.  The CCCP clusters considered in this sample were
observed using CFHT Megacam as part of an optical follow-up of X-ray
selected galaxy clusters (e.g. Bildfell et al. 2008). The processing
of the CFHTLS wide data plus the northern extension and of the CCCP
Megacam data are described in Hoekstra et al. (2006) and Bildfell et
al. (2008) respectively.  Image data is available in the $g$ and $r$
bands.  Table \ref{tab_megacam} describes the main characteristics of
each data set with Figure \ref{fig_comp} showing $r$-band number
counts for the CCCP and W1 data compared with the CFHT Deep survey.
We adopt a magnitude limit $r=23.5$ in all subsequent analyses in
order that photometric samples drawn from all three data sets can be
reasonably considered to be complete.  Given the overlap in the
photometric filters employed by XMM-LSS and CCCP we consider the
photometric properties of cluster galaxies in the combined sample as
determined using $g$ and $r$ bands.

\begin{table}
\centering
\caption{Characteristics of the optical data.}
\label{tab_megacam}
\begin{tabular}{lcccc}
\hline
Sample & $g$-band & \multicolumn{2}{c}{$r$-band} \\
       & $t_{exp}$(s) & $t_{exp}$(s) & seeing (\arcsec) \\
\hline \hline
XMM-LSS & 2500 & 2000 & 0.8 \\
CCCP & 1800 & 4800 & 0.7 \\
\hline
\end{tabular}
\end{table}

Source extraction and photometry were performed using {\tt SExtractor
  v2.5.0} (Bertin \& Arnouts 1996).  Image regions affected by
saturated stars and detector artefacts were excluded.  Zero point
information for sources detected in the CFHTLS W1 area plus northern
extension was extrapolated from common sources detected in the Sloan
Digital Sky Survey equatorial patch which overlaps the southern edge
of the W1 area.  Source photometry is quoted in AB
magnitudes measured within 3\arcsec\ diameter circular apertures.

\begin{figure}
\includegraphics[width=84mm]{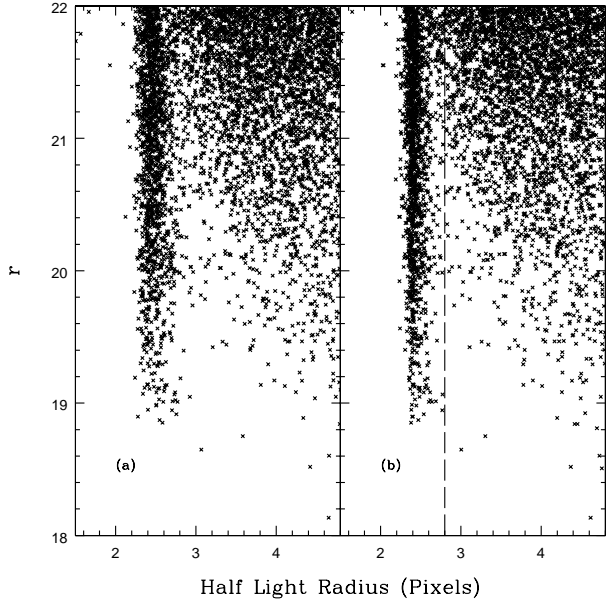}
\caption{Location of the Stellar Locus in a typical CFHTLS Wide
  Megacam field.  (a) Before PSF scaling is applied.  (b) After PSF
  scaling is applied. See text for more details.}
\label{fig_stargal}
\end{figure}

Star-galaxy separation was performed by considering the $r$-band
half-light radius (HLR) versus $r$-band magnitude distribution for
sources in each Megacam field (Figure \ref{fig_stargal}).  The
properties of instrument point spread function (PSF) that determines
the half-light radius of the stellar locus varies systematically over
the Megacam field.  Considered over the entire field, this variation
broadens the stellar locus and reduces the effectiveness of a single
HLR cut at excluding stellar sources.  The field itself is mapped onto
a $4\times9$ array of CCD detectors.  We therefore determined the
weighted mean HLR of the stellar locus in each Megacam CCD and
rescaled the HLR values of all sources in each CCD such that the
stellar values mapped on to a single locus defined in a reference CCD
close to the optical centre of the detector.  This operation reduced
the effective dispersion of sources defining the stellar locus in each
field (Figure \ref{fig_stargal}) and permitted the application of a
single threshold to exclude stellar sources\footnote{We note that a
  component of the PSF variation over the Megacam field is
  anisotropic. However, for the purpose of star-galaxy separation,
  application of an isotropic scaling factor is sufficient.}.

\section{Colour Magnitude Diagrams}
\label{sec_cmd}

Colour magnitude diagrams for each of the hot, mid and cool cluster
samples are displayed in Figures \ref{fig_cmd_hot}, \ref{fig_cmd_mid}
and \ref{fig_cmd_cool} respectively. All sources lie within $r_{500}$
of the measured X-ray centre of each cluster.

Computation of the blue fraction in each cluster first requires the
red sequence relation defining the linear colour sequence followed by
cluster early-type galaxies to be determined. This provides a
reference value relative to which the blue galaxy population in each
cluster may be defined. To determine the location of the red sequence
we first applied a statistical background subtraction procedure to
each cluster to highlight the population of cluster members.  We
applied the method of \cite{pimb02} whereby the field
population and the cluster plus field population were each represented
on a colour-magnitude grid. The probability that a galaxy occupying a
particular grid position is field galaxy is then
\begin{equation}
{
P(col,mag)_{Field}=\frac{A \times N(col,mag)_{Field}}{N(col,mag)_{Cluster+Field}},
}
\label{eqn_prob}
\end{equation}
were $A$ is an areal scaling factor used to match the area of the
field sample to the cluster area. The field colur magnitude
distribution is computed employing all galaxy sources at radii $>8
 \times r_{500}$ from the X-ray cluster centre. For each galaxy within
$r_{500}$ of the cluster centre, membership was determined by
comparing a random number in the interval [0,1] to the field
probability value.  This procedure was repeated 100 times for each
cluster.

\begin{figure}
\includegraphics[width=84mm]{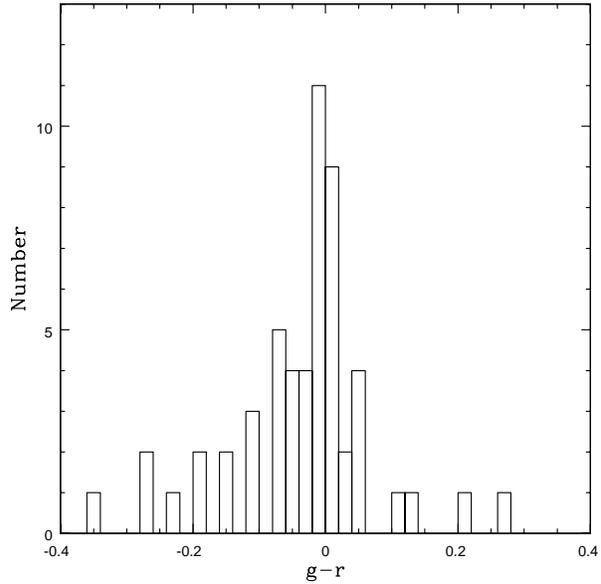}
\caption{Red edge diagram for XLSSC 22. See text for more details.}
\label{fig_rededge}
\end{figure}

\begin{figure*}
\begin{minipage}{126mm}
\begin{center}
\includegraphics[width=150mm]{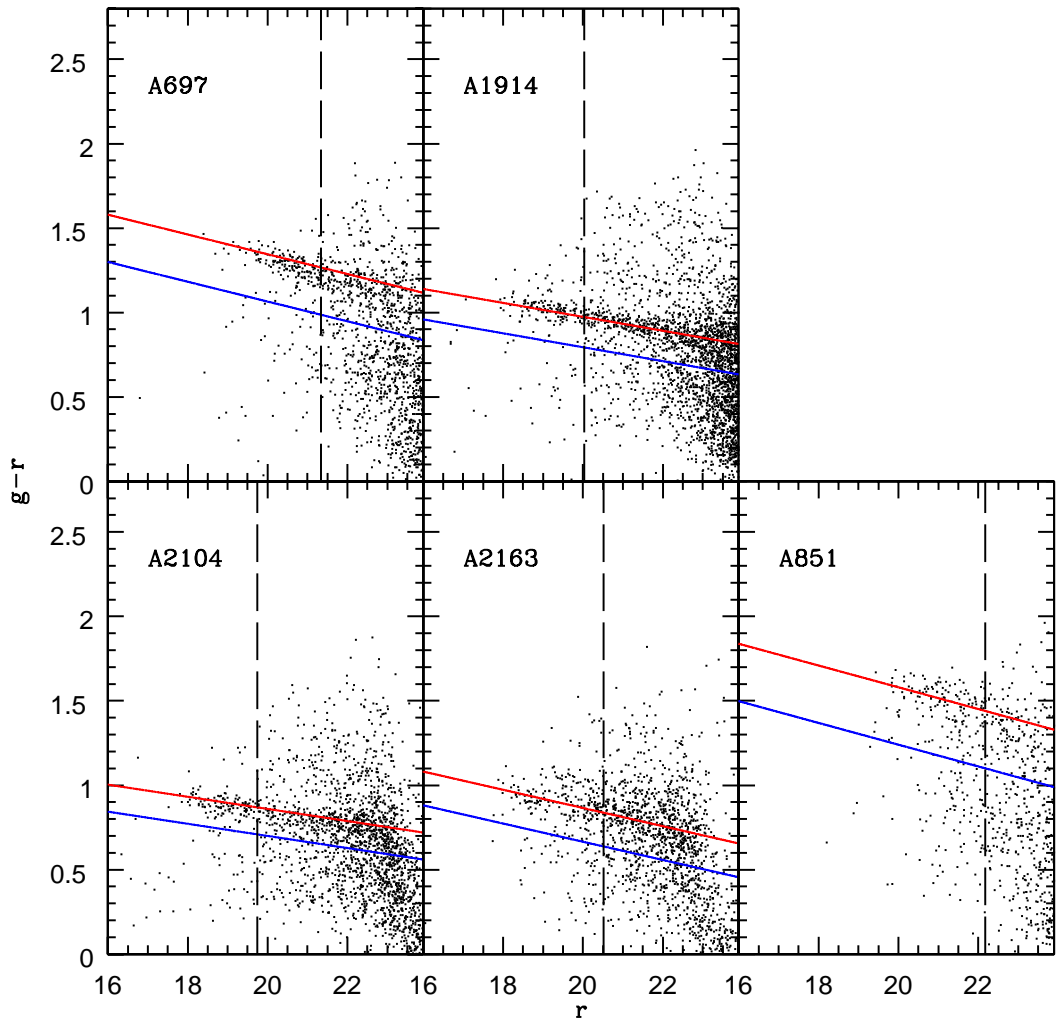}
\caption{Hot sample clusters. All sources within $r_{500}$ of the
  cluster centre are plotted.  The red line marks the location of the
  red sequence and the blue line marks the Butcher \& Oemler (1984)
  cut as described in the text.  The vertical dashed line indicates
  the $r$ magnitude corresponding to $M_V=-20$ at the cluster
  redshift.}
\label{fig_cmd_hot}
\end{center}
\end{minipage}
\end{figure*}

\begin{figure*}
\begin{minipage}{126mm}
\begin{center}
\includegraphics[width=150mm]{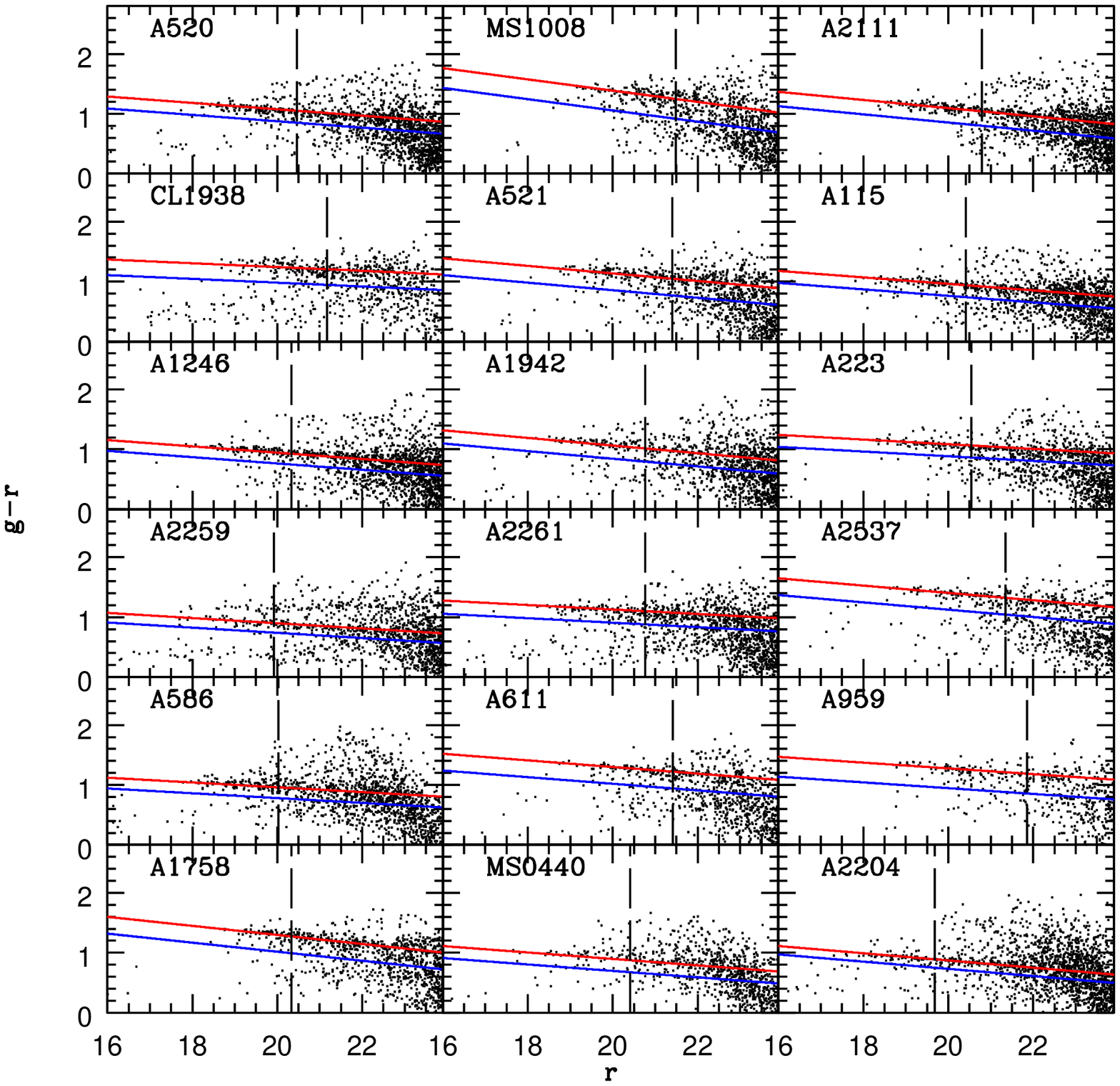}
\caption{Mid sample clusters. All sources within $r_{500}$ of the
  cluster centre are plotted.  The red line marks the location of the
  red sequence and the blue line marks the Butcher \& Oemler (1984)
  cut as described in the text.  The vertical dashed line indicates
  the $r$ magnitude corresponding to $M_V=-20$ at the cluster
  redshift.}
\label{fig_cmd_mid}
\end{center}
\end{minipage}
\end{figure*}

\begin{figure*}
\begin{minipage}{126mm}
\begin{center}
\includegraphics[width=150mm]{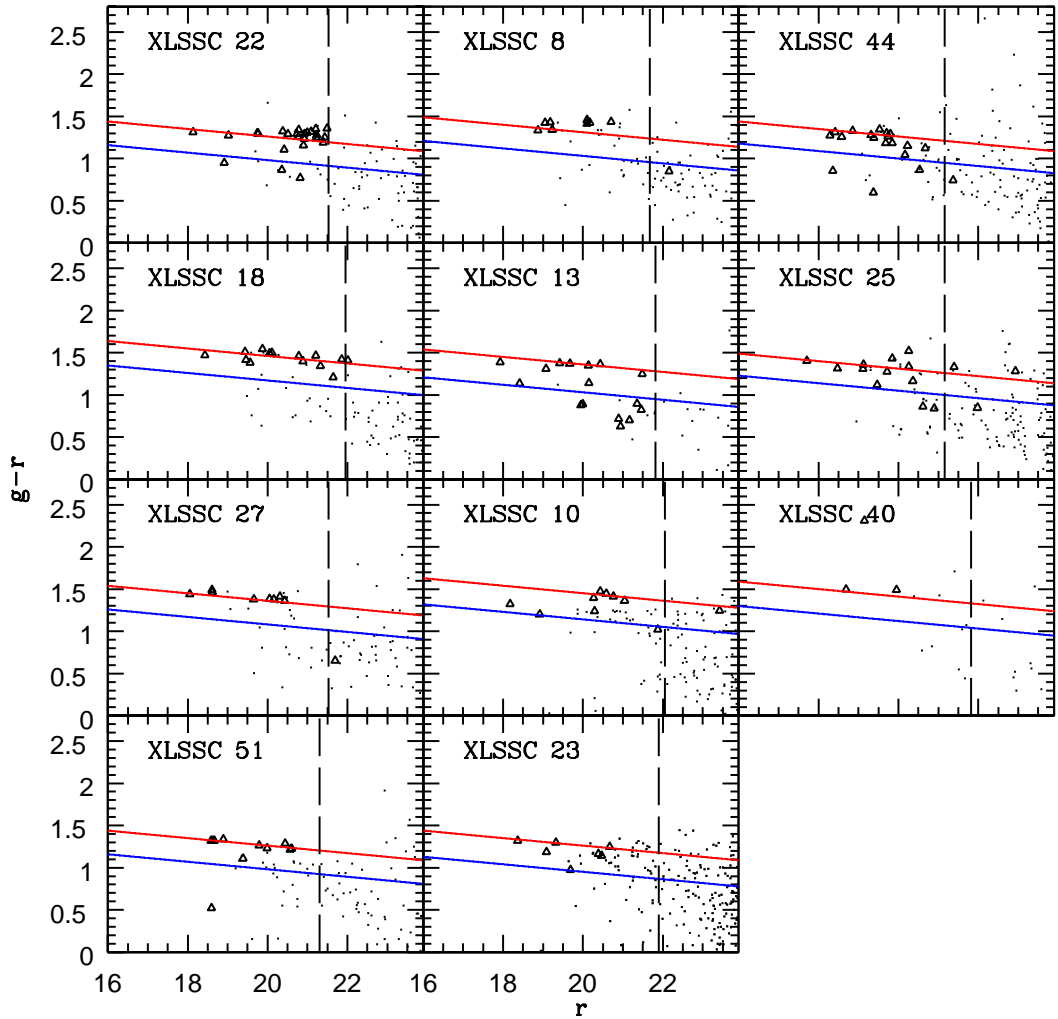}
\caption{Cool sample clusters. All sources within $r_{500}$ of the
  cluster centre are plotted.  The red line marks the location of the
  red sequence and the blue line marks the Butcher \& Oemler (1984)
  cut as described in the text.  The vertical dashed line indicates
  the $r$ magnitude corresponding to $M_V=-20$ at the cluster
  redshift. The triangles represent spectroscopically confirmed
  members.}
\label{fig_cmd_cool}
\end{center}
\end{minipage}
\end{figure*}

For each realisation, the red sequence location was computed employing
a weighted, linear least squares fit and the mean slope and zero-point
was calculated from the distribution of 100 values.  To check and, if
necessary, refine the location of the mean red sequence for each
cluster we considered the ``red edge'' diagram for each system (an
example of which is displayed in Figure \ref{fig_rededge}).  The red
edge diagram shows the number of galaxies located at a colour offset
$\Delta(g-r)$ at fixed $r$-magnitude from the red sequence in a
particular cluster.  The zero-point of the red sequence was then
adjusted to set the peak in the red edge distribution at zero colour
offset.

This procedure worked well for the mid and hot cluster
samples. However, following statistical background subtraction, the
cool clusters typically possess insufficient cluster members to
generate a reliable red sequence fit.  In this case the red sequence
relation computed for the stacked mid cluster sample described in
Section \ref{sec_stack} was applied to the cool clusters.  The red
sequence zero-point was adjusted using the red edge diagram procedure
to provide an improved fit to individual clusters.

\begin{figure}
\includegraphics[width=84mm]{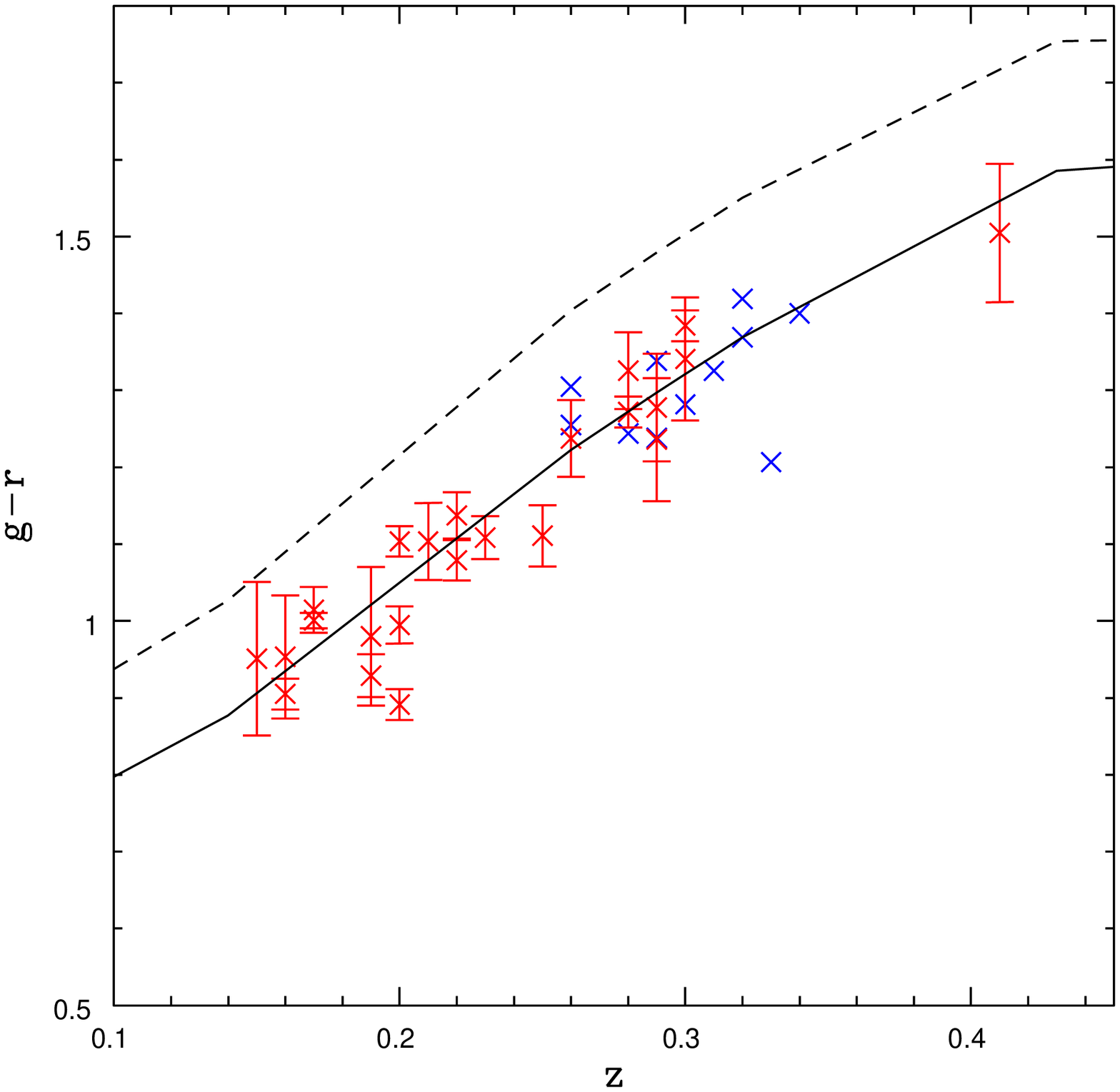}
\caption{The red crosses indicate the $g-r$ colour of the fitted red
  sequence relation measured at $M_{V}=-20$ for the mid and hot
  samples.  The dashed line marks the 100\% elliptical model for
  reference. Blue crosses indicate the location of the red sequence in
  the cool clusters determined using the mid cluster red sequence
  relation (see text for more details).}
\label{fig_gr}
\end{figure}

Figure \ref{fig_gr} indicates the $g-r$ colour of the red sequence
measured at fixed absolute magnitude ($M_V = -21$) versus redshift for
the mid and hot clusters. The colour uncertainty is computed as the
standard deviation of the red sequence zero-points as measured for the
100 realisations of the cluster subtraction procedure.  The
$k$-correction employed to convert absolute $V$-magnitude to apparent
$r$-magnitude at the redshift of each cluster assumes a early-type
galaxy spectral energy distribution (SED; Kinney et al. 1996).  We
note that the exact choice of template is not a significant factor
when computing the apparent $r$-magnitude reference location on the
red sequence as the $k$-correction is dominated by the bandwidth term
at $z\sim 0.3$. Furthermore, due to the small slope of the fitted red
sequence relation in each case ($\sim 0.05$) small systematic
magnitude errors result in negligible colour uncertainties.

The best fitting SED describing the observed colour evolution of the
mid and hot cluster red sequences is computed employing a linear
interpolation between an early-type (Ell) and early-type spiral (Sab)
SED, e.g.
\begin{equation}
{
{\rm SED}(\lambda)=(1-x)~{\rm Ell}(\lambda)+x~{\rm Sab}(\lambda),
}
\label{eqn_sed}
\end{equation}
where the parameter $x$ is computed using a $\chi^2$ minimisation
procedure. The best fitting SED model was found to be approximately
70\% Elliptical and 30\% Sab. We employ this template as a reference
point from which to determine accurate $k$-corrections and blue galaxy
colour thresholds for all clusters (cool, mid and hot) in the
following discussion.

\section{Blue Fractions}

The blue fraction of each cluster was computed following the
definition of Butcher \& Oemler (1984). Following their approach, all
galaxies displaying an absolute magnitude $M_{V}\le -20$ are
considered and blue galaxies are defined as those displaying a rest
frame colour offset $\Delta(B-V)=-0.2$ measured relative to the red
sequence.  We add the further criterion that galaxies within $r_{500}$
of the cluster X-ray centre are considered part of the ``total'',
i.e. cluster plus field, populations, and galaxies at clustercentric
radii $> 8 \times r_{500}$ and within the same Megacam field are
considered as the ``field'' population. The blue fraction within each
cluster is then computed as
\begin{equation}
{
f_{B}=\frac{N_{Blue,Total}-A~N_{Blue,Field}}{N_{Total}-A~N_{Field}},
}
\label{eqn_bfrac}
\end{equation}
where $N_{Blue,Total}$ is the number of blue galaxies in the cluster
plus field, $N_{Blue,Field}$ is the number of blue field galaxies,
$N_{Total}$ is the total number of galaxies in the cluster plus field
and $N_{Field}$ is the total number of field galaxies. The symbol $A$
denotes an areal scaling factor to correct the field population
area to that of the cluster area.

The rest frame Butcher \& Oemler (1984) magnitude and colour criteria
were expressed as observed frame $r$-magnitude and $\Delta(g-r)$
colour offsets at the redshift of each cluster by considering the
galaxy SED implied by the cluster red sequence colour (Section
\ref{sec_cmd}). The colour offset was computed by generating a second
SED model according to Equation \ref{eqn_sed} displaying
$\Delta(B-V)=-0.2$ compared to the reference SED describing the
cluster red sequence relation. This ``blue-cut'' SED model was found
to be 75\% Sab and 25\% Sbc. Employing this SED model
$\Delta(g-r)$ corresponding to $\Delta(B-V)=-0.2$ was calculated for
each cluster at the appropriate redshift.

The blue fraction error is estimated by computing a distribution of
blue fraction values for each cluster.  Each blue fraction value is
computed as follows: circular apertures of radius $r_{500}$ are placed
at random locations within the Megacam field of each cluster. The
galaxy population within each aperture is employed as the background
value for Equation \ref{eqn_bfrac}.  The error on the blue fraction
for each cluster is then estimated as the interval about the median
blue fraction value containing 67\% of the distribution.  Errors
computed using this method are typically 1.7 times larger than those
computed assuming Poissonian uncertainties alone.

\begin{figure}
\includegraphics[width=84mm]{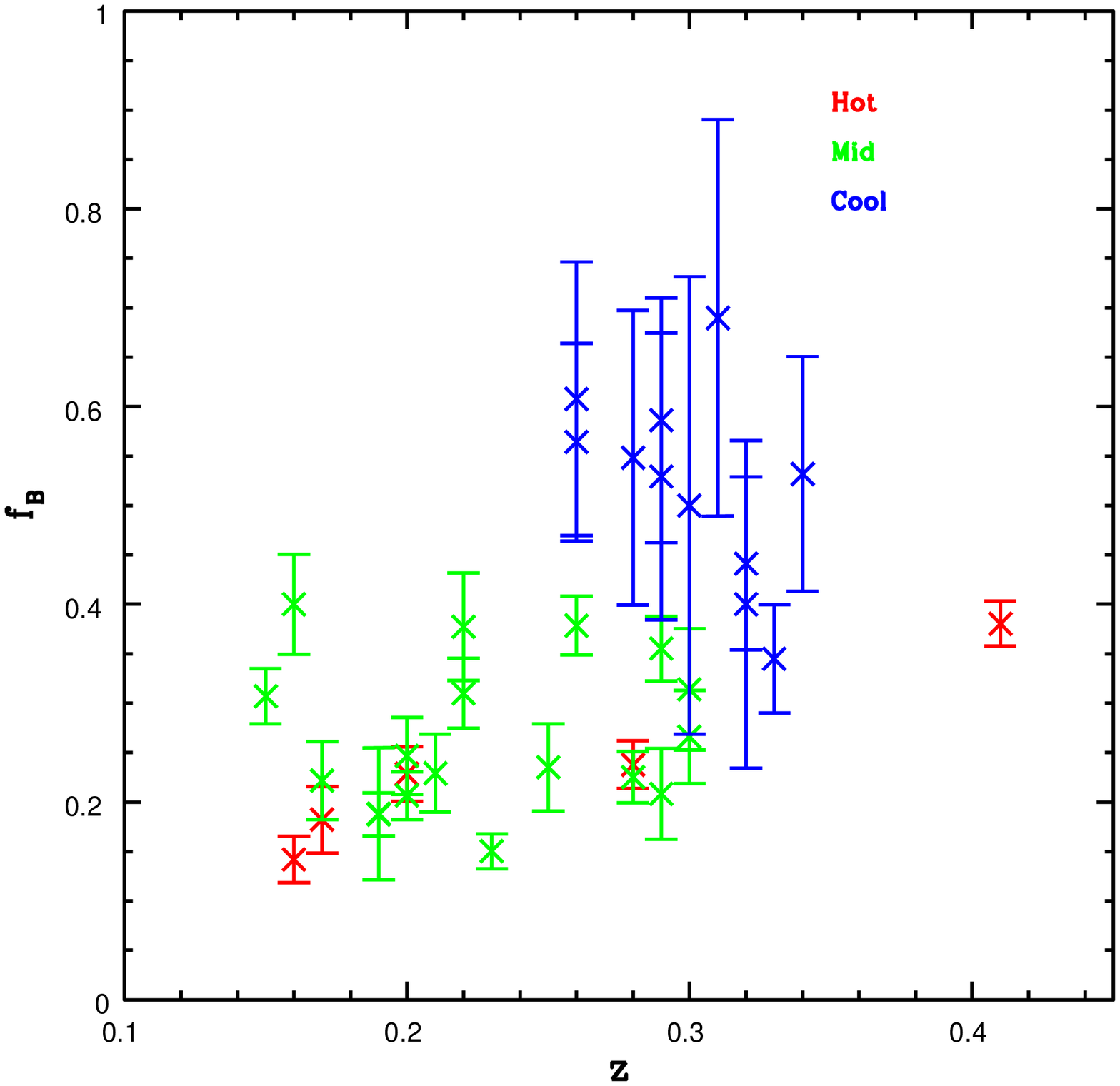}
\caption{Cluster blue fractions as a function of redshift.}
\label{fig_red}
\end{figure}

Figure \ref{fig_red} displays the blue fraction as a function of
redshift for all clusters and shows an apparent trend to observe
increasing blue fraction versus redshift.  However, splitting the
sample by X-ray temperature reveals that this trend may instead arise
from the varying global environment of each cluster modulo the
slightly different redshift interval covered by each of the cool, mid
and hot samples.

\begin{figure}
\includegraphics[width=84mm]{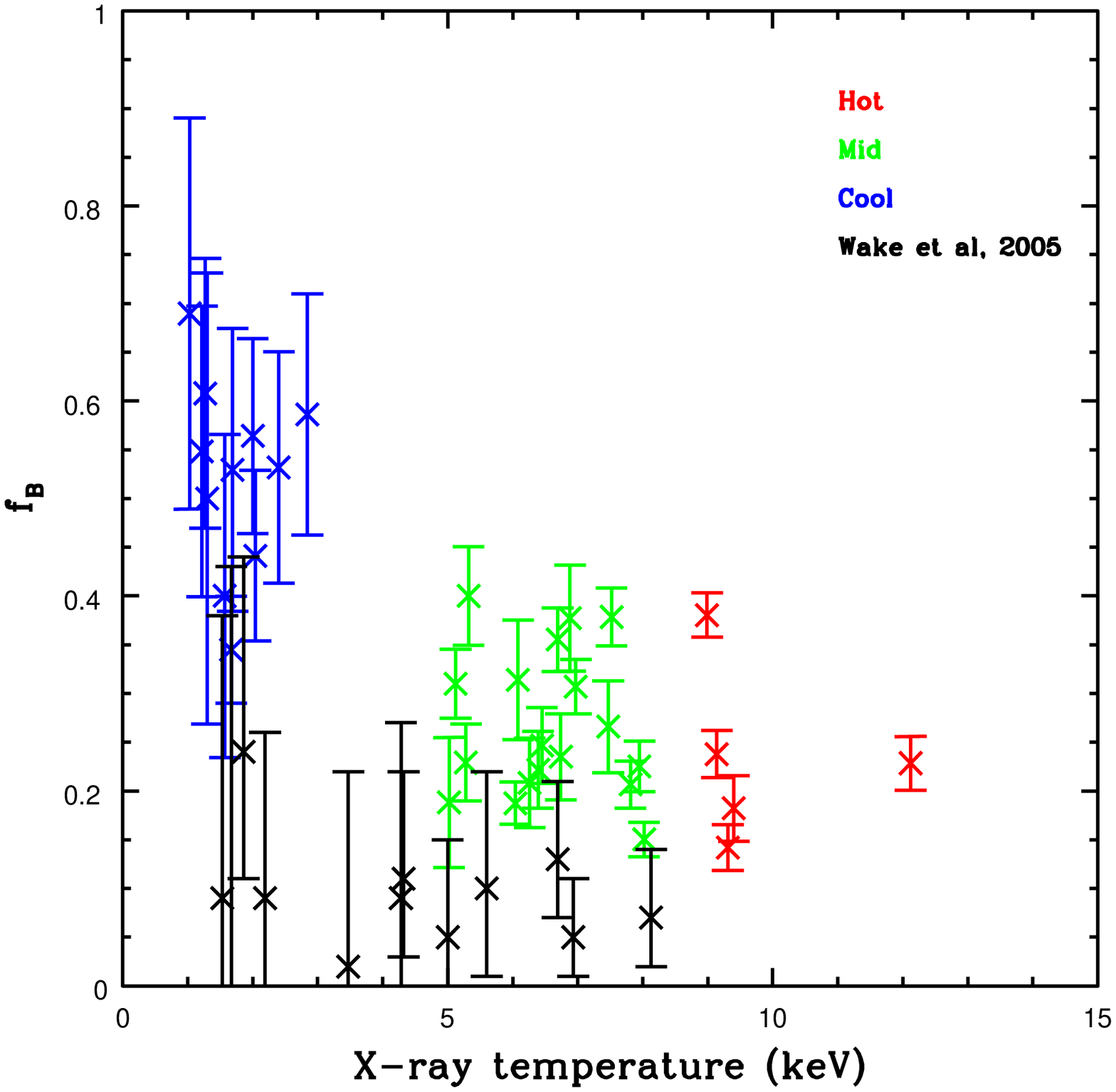}
\caption{Cluster blue fraction as a function of X-Ray temperature.}
\label{fig_temp}
\end{figure}

Figure \ref{fig_temp} displays the blue fraction as a function of
cluster X-ray temperature for the cool, mid and hot samples. In
addition, blue fraction values and Poisson uncertainties are listed in
Table \ref{tab_bfrac}. The data indicate that typical blue fraction in
each cluster and the dispersion in blue fraction values within a given
temperature sub-sample increase as the temperature of the X-ray
cluster decreases. 

Clusters of similar properties have been studied by \cite{wake05} and
we also compare their results to ours as a function of cluster X-ray
temperature in Figure \ref{fig_temp}.  One immediately notes that the
blue fraction values for the current sample are consistently larger
than the sample of Wake et al. though nominally covering a similar
range of X-ray temperature.  Blue fractions in the Wake et al. sample
are computed within an aperture of radius one-third of the virial
radius. This corresponds to a radius approximately equal to
$0.7~r_{500}$ and we have corrected the Wake et al. values to the
aperture used in the current paper using the blue fraction versus
radius curve appropriate to each cluster temperature (shown in Figure
\ref{fig_radvar}).  This correction is necessarily approximate and
clearly offsets between the two samples remain.  The remaining
differences arise from the different methods used to determine blue
fraction values in each sample.  Specifically, Wake et al. assume that
the SED representing the cluster colour magnitude relation is a pure
elliptical model rather than the composite elliptical plus Sab model
adopted in the current paper.  The use of different spectral models
results in different $k$--corrections versus redshift with the
consequence that the observed frame colour offset employed to define
to the blue cut in each cluster is typically larger in the Wake et
al. sample than that employed here. The application of a bluer cut
relative the CMR in a given cluster results in a smaller blue
fraction.  We therefore conclude that the remaining offset in blue
fraction versus temperature between the Wake et al. sample and that
presented in the current paper results from such methodology
differences.

\begin{figure}
\includegraphics[width=84mm]{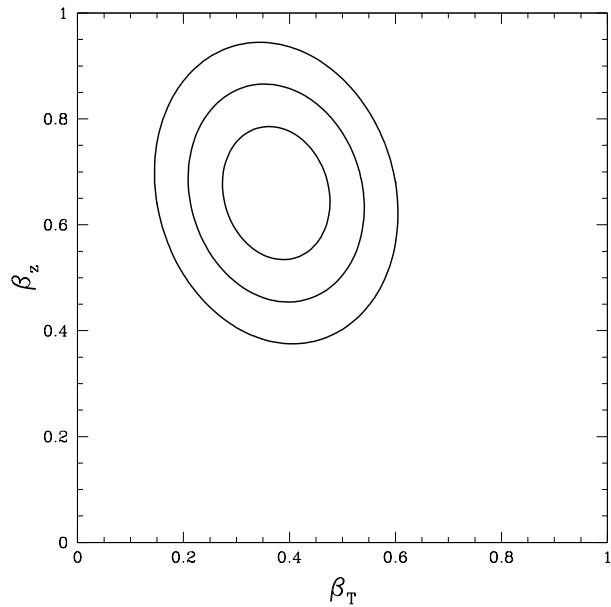}
\caption{Confidence intervals on the fitted values of $\beta_z$ and
  $\beta_T$ (1, 2, and 3-sigma confidence intervals are shown).}
\label{fig_cont}
\end{figure}

As noted by \cite{marg01} and \cite{hansen09}, the observed trends in
cluster blue fractions can be explained as a function of both redshift
and cluster mass. In order to determine the relative influence of
varying redshift and cluster temperature (here used as a proxy for
cluster mass) upon the blue fractions we fit the data for the three
samples with a simple function of the form
\begin{equation}
{
f_B(z,T) = \beta_0 + \beta_z z + \beta_T T^{-1}
}
\label{eqn_fit}
\end{equation}
where $\beta_0$, $\beta_z$ and $\beta_T$ are constants to be
determined employing a minimum $\chi^2$ algorithm and $T$ is expressed
in keV.  The best fitting values are $\beta_0 = 0.04 \pm 0.02$,
$\beta_z = 0.67 \pm 0.08$ and $\beta_T = 0.42 \pm 0.07$.  Confidence
intervals of the fitted values of $\beta_z$ and $\beta_T$ are shown in
Figure \ref{fig_cont}.  The minimum value of $\chi^2$ obtained using
this procedure is approximately 3 per degree of freedom and, if the
blue fraction errors are accurate, may indicate the presence of
intrinsic scatter in the distribution of blue fraction values. The
current sample of XMM-LSS and CCCP clusters is insufficiently large
and does not cover a large enough interval in either redshift or
temperature to constrain both the exponents of the redshift and
temperature dependence of Equation \ref{eqn_fit} in addition to the
coefficients. We can rule out no redshift evolution but the
temperature dependence is relatively unconstrained (and is complicated
by the degeneracy with $\beta_0$ as the power law exponent of $T$
tends to zero).

An alternative approach is to investigate the trends in the data
employing a non-parametric approach. In this case we apply a partial
Spearman rank analysis to the data in order to examine the extent of
any correlation between $f_B$ and either temperature or redshift while
controlling for the variation of the second variable.  The partial
correlation coefficient describing blue fraction and redshift while
controlling for temperature is $r(f_B,z,T) = 0.39$ while the
corresponding coefficient describing blue fraction versus temperature
while controlling for redshift is $r(f_B,T,z) = -0.64$.  The
probability of obtaining $r=-0.64$ from the sample by chance is
approximately 0.2\% whereas the corresponding probability for the
value $r=0.39$ is somewhat greater at approximately 1.8\%.  Overall,
both the parametric and non-parametric approach confirm the evolution
in the cluster blue fraction with both redshift and cluster
temperature.  The significance of the greater blue fraction versus
decreasing temperature can also be seen in Figure \ref{fig_fred} where
the average blue fraction for the cool sample is significantly higher
than either the blue fraction for the mid and hot samples.

\begin{figure}
\includegraphics[width=84mm]{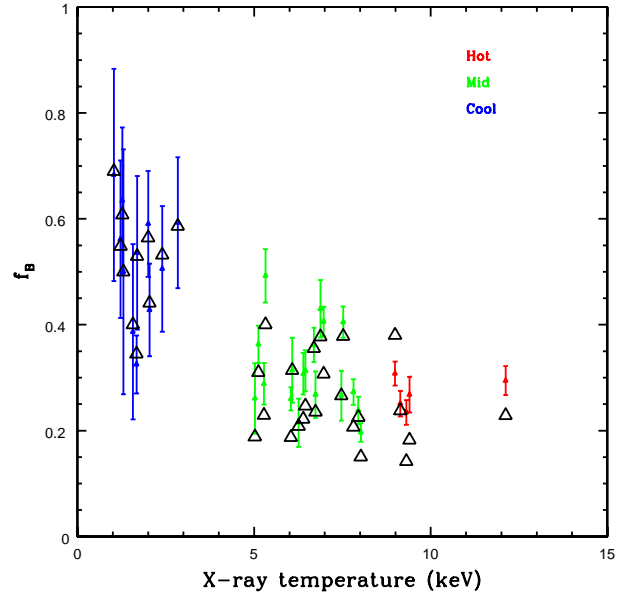}
\caption{Cluster blue fraction as a function of X-Ray temperature. The
  black points indicate the result of correcting the original blue
  fraction values to a common epoch at $z=0.3$.}
\label{fig_noz}
\end{figure}

One may further investigate the dependence of blue fraction on cluster
temperature by correcting $f_B$ values for individual clusters to a
common epoch at $z=0.3$ using Equation \ref{eqn_fit} and the
best-fitting coefficients. The corrected $f_B$ values are displayed in
Figure \ref{fig_noz} and confirm the trend to observe greater blue
fraction values in cooler clusters.
We consider further the implications of this trend in Section
\ref{sec_conc}. We next consider the effects of the various choices
made during the blue fraction analysis on the overall robustness of
the blue fraction trends versus redshift and temperature.

\subsection{Testing the blue fraction computation assumptions}

\begin{figure}
\includegraphics[width=84mm]{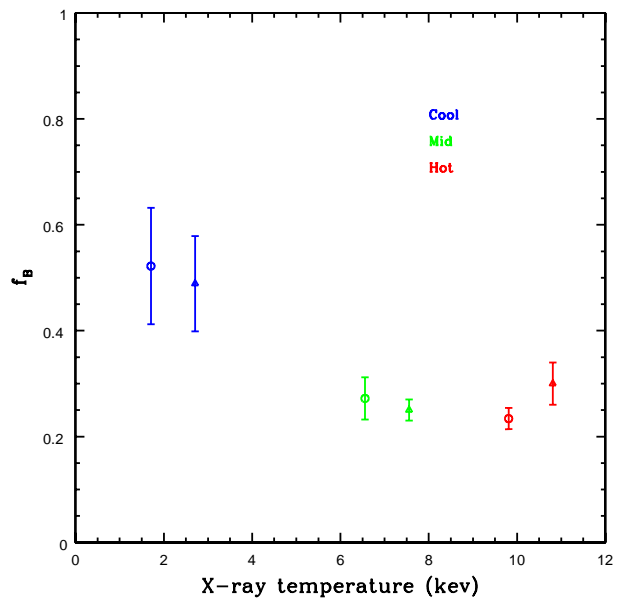}
\caption{A comparison of blue fraction computation methods versus
  temperature. The circular points indicate the average blue fraction
  in each of the three temperature sub-samples computed using the BO84
  method. The triangles indicate the average blue fraction per
  temperature sub-sample computed using the definition $f_B^\prime =
  1-f_R$ (see text for details). The triangles have been offset in
  temperature from the circles for clarity }
\label{fig_fred}
\end{figure}

The cluster blue fraction may be computed employing one of a number of
crtieria to segregate the red and blue cluster galaxy populations.  We
apply the original definition of \cite{bo84} in this paper but note
the critiscisms of \cite{andreon05} who argue that adopting a fixed
rest frame colour offset relative to the red sequence is prone to give
misleading blue fraction trends as galaxies of different SEDs drift in
and out of the adopted colour interval as a funtion of redshift.  To
assess whether this is an important consideration in our analysis we
compare the average blue fraction per cluster sub-sample to a modified
blue fraction determined by fitting the red fraction associated with
the stacked CMD of clusters by temperature.  We follow the method
described by \cite{loh08} and model the red wing of the red sequence
of the CMD of all clusters stacked by temperature once $k$-corrected
to a common redshift $z=0.3$ (we defer the description of the stacking
procedure to Section \ref{sec_stack}). The red wing is modelled as a
double Gaussian function and the best-fitting function is reflected
about the location of the red sequence to determine the number of red
galaxies in each distribution. We then compute the red fraction in an
analogous manner to the blue fraction and define a modified blue
fraction as $f_B^\prime = 1-f_R$.  We compare the original and
modified blue fraction values for each temperature sub-sample in
Figure \ref{fig_fred} and note that each approach reveals the same
trend (and that the points are identical within the errors).  We are
therefore satisfied that the original BO84 definition produces
reliable results and we avoid the requirement to fit double Gaussian
features to individual cluster colour distributions which, in the case
of the cool clusters, typically result in poorly constrained models.

\begin{table}
 \caption{Blue Fractions for all clusters employing a limiting magnitude
$M_{V}=-20$ within $r_{500}$ of the cluster centre.}
\label{tab_bfrac}
\begin{tabular}{lcc}
\hline
Cluster & $f_{B}$  & Number of blue galaxies\\
\hline

XLSSC 13 & $ 0.69 \pm 0.20 $ & 11\\
XLSSC 51 & $ 0.55 \pm 0.15 $ & 11\\
XLSSC 44 & $ 0.61 \pm 0.14 $ & 13\\
XLSSC 08 & $ 0.50 \pm 0.23 $ & 14\\
XLSCC 40 & $ 0.40 \pm 0.17 $ & 5\\
XLSSC 23 & $ 0.35 \pm 0.05 $ & 16\\
XLSSC 22 & $ 0.53 \pm 0.15 $ & 10\\
XLSSC 25 & $ 0.56 \pm 0.10 $ & 17\\
XLSSC 18 & $ 0.44 \pm 0.09 $ & 14\\
XLSSC 27 & $ 0.59 \pm 0.12 $ & 12\\
XLSSC 10 & $ 0.53 \pm 0.12 $ & 20\\ \hline

MS0440+02 & $ 0.19 \pm 0.07 $ &16\\
A1942 & $ 0.31 \pm 0.04 $ & 49\\
A0223 & $ 0.23 \pm 0.04 $ & 30\\
A2259 & $ 0.40 \pm 0.05 $ & 50\\
A1246 & $ 0.24 \pm 0.02 $ & 30\\
A2537 & $ 0.31 \pm 0.06 $ & 54\\
A0959 & $ 0.21 \pm 0.05 $ & 25\\
A0586 & $ 0.22 \pm 0.04 $ & 39\\
A0115 & $ 0.25 \pm 0.04 $ & 38\\
A0611 & $ 0.36 \pm 0.03 $ & 65\\
A0521 & $ 0.24 \pm 0.05 $ & 55\\
A2261 & $ 0.38 \pm 0.05 $ & 80\\
A2204 & $ 0.31 \pm 0.03 $ & 35\\
MS1008-12 & $ 0.27 \pm 0.05 $ & 58 \\
CL1938+54 & $ 0.38 \pm 0.03 $ & 95\\
A0520 & $ 0.21 \pm 0.02 $ & 43\\
A1758 & $ 0.23 \pm 0.03 $ & 52\\
A2111 & $ 0.15 \pm 0.02 $ & 26\\ \hline

A0851 & $ 0.38 \pm 0.02 $ & 97\\
A0697 & $ 0.24 \pm 0.02 $ & 54\\
A2104 & $ 0.14 \pm 0.02 $ & 24\\
A1914 & $ 0.18 \pm 0.03 $ & 35\\
A2163 & $ 0.23 \pm 0.03 $ & 69\\

\hline
\end{tabular}
\end{table}

\begin{figure}
\includegraphics[width=84mm]{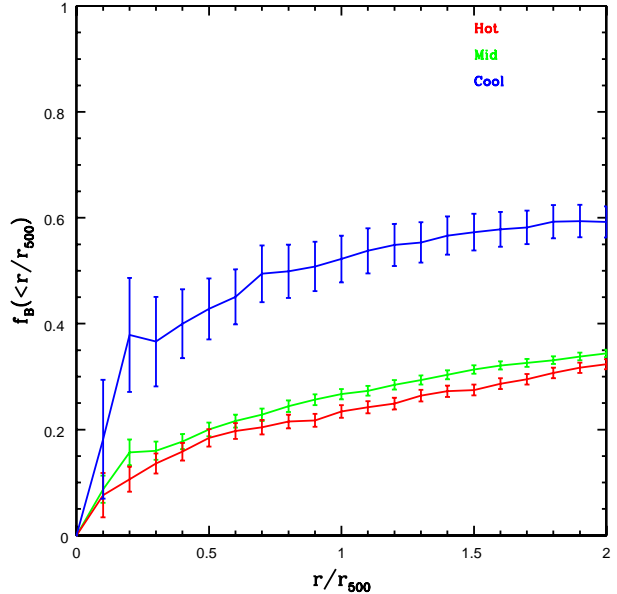}
\caption{Radial variation of the cluster blue fraction}
\label{fig_radvar}
\end{figure}

We next consider the variation of the computed blue fraction with
assumed cluster radius to a) investigate whether the choice of radius
introduces a bias into the trend of observed blue fraction versus
temperature, b) determine the effect of uncertainty in the estimated
$r_{500}$ value for each cluster on $f_B$ and c) relate blue fractions
computed for the current sample to comparable studies in the
literature. Figure \ref{fig_radvar} displays the average blue fraction
per temperature sub-sample computed within circular apertures of
radius expressed as a fraction of $r_{500}$.  This variation of the
blue fraction with characteristic radius has been noted in previous
studies and the trend displayed in Figure \ref{fig_radvar} are
consistent with those of \cite{fairley01}, \cite{ellingson01} and
\cite{wake05}.  Though we discuss this trend further in Section
\ref{sec_conc} we note here that the increasing trend of blue fraction
with scaled aperture radius is nominally consistent with a blue
infalling field population that is processed by ram pressure stripping
upon falling into each cluster. However, the details of this
conclusion will be discussed in Section \ref{sec_conc}.  The average
blue fraction at $r_{500}$ in each temperature sub-sample reflects the
trend observed in Figure \ref{fig_temp}.  What can be noted at this
point is that the exact choice of radius wihtin which blue fractions
are computed does not affect the conclusion that cooler clusters
display greater blue fractions than hotter clusters (within the range
$0.5 ~r_{500} < r < 2 ~r_{500}$ investigated here)\footnote{Furthermore,
  we note that the value of $r_{500}$ in each case is computed
  employing the observed cluster temperature and an assumed
  mass-temperature relation. X-ray spectral measurements from which
  the cluster temperature is computed are typically extracted within
  an on-sky aperture of size of the order of $r_{500}$ in each
  cluster. Therefore, employing computed $r_{500}$ values to define
  aperture sizes up to several times the virial radius would invlove a
  considerable extrapolation beyond the scale on which the X-rays are
  measured.}.

Uncertainty in the temperature computed for each cluster propagates to
an error in the value of $r_{500}$ computed for each cluster. The
typical temperature error for the XMM-LSS clusters contributing to the
cool sample are of the order of 10-20\% (Willis et al. 2005).  The
gradient of the trend of $f_B$ versus $r/r_{500}$ estimate at
$r_{500}$ is approximately 0.15. This indicates that uncertainty in
cluster X-ray temperature will contribute a fractional $f_B$ error of
order a few percent, i.e. small compared to the Poissonian error in
$f_B$ computed for an individual cluster.

The mid temperature CCCP cluster MS1008 is common to both this work
and that of \cite{kodama01} and provides a useful check upon our
results.  The value of the blue fraction obtained by these authors was
$f_{B}=0.161\pm0.036$ and was calculated within $R_{30}$ (the radius
containing 30\% of the total number of galaxies in the cluster)
corresponding to an angular radius of 2.82\arcmin.  We obtain a value
$f_{B}=0.27\pm0.04$ within $r_{500}$ corresponding to an angular
radius of 4.44\arcmin.  The radial cut used by \cite{kodama01}
corresponds to approximately $0.63~r_{500}$ and when the blue fraction
was recalculated within this fraction of $r_{500}$, a value of $f_{B}=
0.18\pm0.04$ was obtained, improving the agreement with
\cite{kodama01}.
We note that the errors
quoted here are purely Poissonian and do not account for the effect
that minor differences in the methodology used in each study
(e.g. $k$-correction, computation of the red sequence relation) have
upon the computed value of $f_B$.

\begin{figure}
\includegraphics[width=84mm]{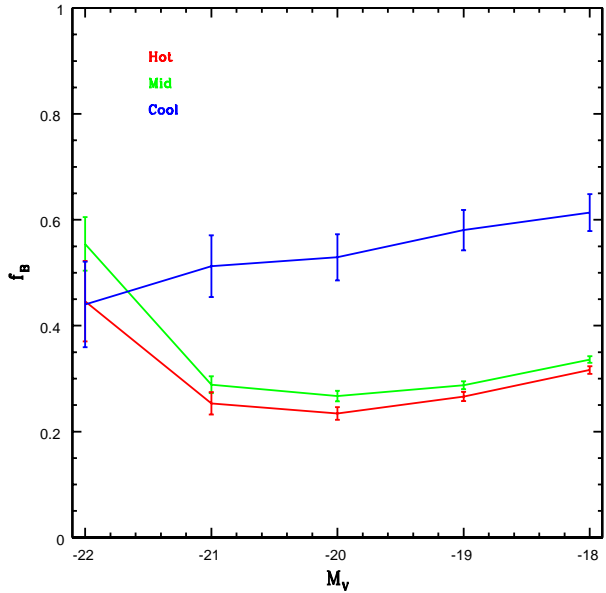}
\caption{Variation of the cluster blue fraction for each cluster
  sample with faint magnitude cut.}
\label{fig_magvar}
\end{figure}

We next consider the effect of extending the faint absolute magnitude
cut applied when computing the blue fraction.  Figure \ref{fig_magvar}
displays the average blue fraction per temperature sub-sample computed
within a circular aperture of $r_{500}$ as the applied faint magnitude
cut is varied from $M_V=-22$ to -18. The main
feature of the diagram is that the excess fraction of blue galaxies
observed in cool clusters compared to the mid and hot samples
continues to increase as the sample magnitude limit is extended to
fainter magnitudes.

When considered together, the variation of blue fraction with both
radius and magnitude indicate that the excess blue galaxy fraction
observed in cool clusters is dominated by the contrubtion of faint
galaxies at large radius. In the following section we attempt to
identify this population directly on the colour-magnitude diagram for
each temperature sub-sample.

\section{Stacked Colour Magnitude Diagrams}
\label{sec_stack}

\begin{figure*}
\begin{minipage}{150mm}
\begin{center}
\includegraphics[width=150mm]{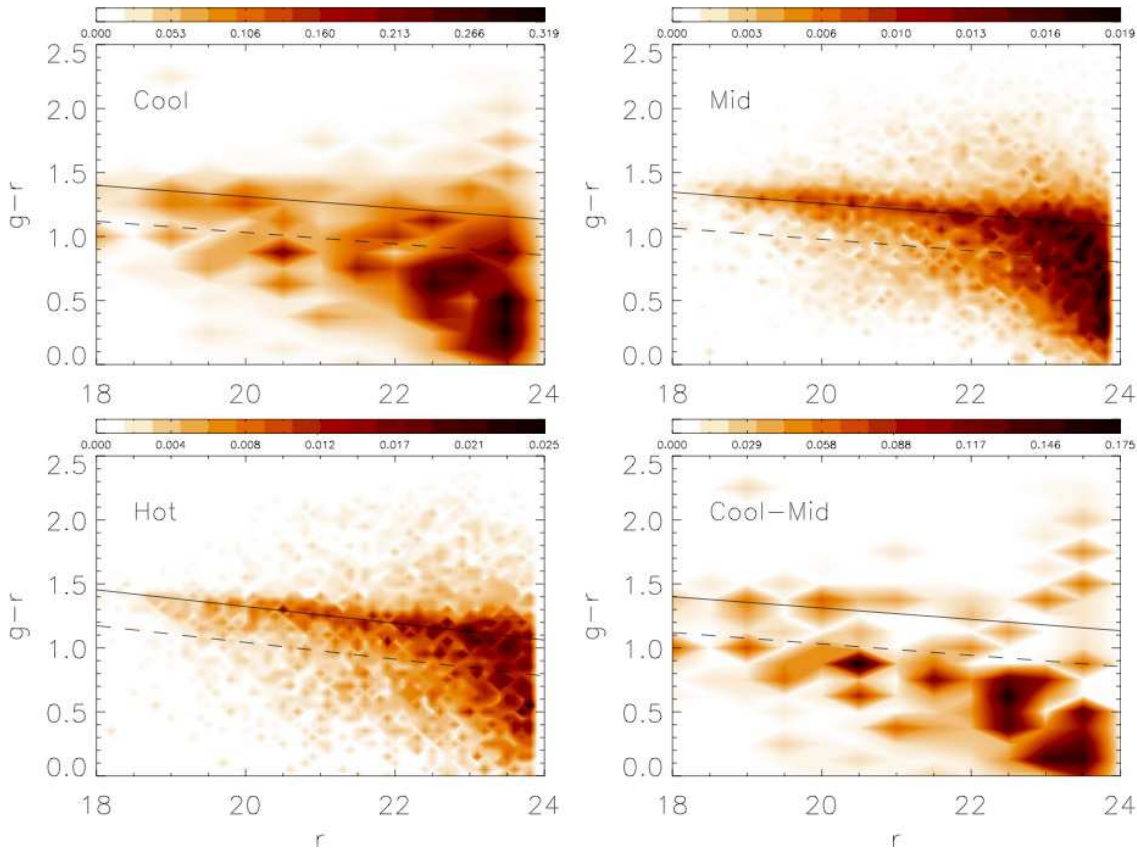}
\caption{Stacked CMDs for the cool, mid, hot and (cool-mid) cluster
  samples (see text for details). All data are $k$-corrected to
  $z=0.3$. In each panel the solid line marks the location of the red
  sequence and the dashed line marks the location of the corresponding
  Butcher-Oemler blue cut.}
\label{fig_stack}
\end{center}
\end{minipage}
\end{figure*}

The previous analyses have identified a relative excess in the
fraction of blue galaxies in cool X-ray clusters compared to the mid
and hot samples.  While the blue fraction is a useful quantity to
investigate, it may prove informative to consider the average CMD for
each temperature sample for which blue fractions are computed.  The
statistical background subtraction method described in Section
\ref{sec_cmd} was applied 100 times to each cluster and the average
CMD was computed by stacking the individual CMDs on a binned colour
magnitude plane in intervals of 1 mag. in magnitude and 0.5 mag. in
colour.  The average background subtracted CMD for each cluster was
then transformed to a common redshift $z=0.3$. The transformation
described the effects of distance dimming and the $k$-correction. The
$k$-correction for each colour pixel on the CMD plane was computed
using the best-fitting interpolated spectral template required to
reproduce the required observed colour value at the cluster redshift.
Individual cluster CMDs within each temperature sub-sample were then
stacked on the $z=0.3$ CMD plane.  Each cluster CMD was assigned equal
weight in the stacking process by normalising all transformed CMDs to
a total contribution of unity summed withing the region $21<r<22$ and
$1<g-r<1.5$.  The stacked colour magnitude diagrams for each sample
are shown in Figure \ref{fig_stack}.  In addition, the fourth panel in
Figure \ref{fig_stack} indicates the result of subtracting the mid
cluster stacked CMD from the corresponding cool cluster sample to
highlight the location on the CMD of the excess fraction of blue
galaxies found in the cool sample. As indicated in the plots, the
relative excess appears uniform across a range of colours bluer than
the red sequence. There is some evidence for an excess of faint
($r> 23$) blue ($g-r\sim0.2$) galaxies in the cool sample. However,
the CFHTLS optical data used to construct the cool CMD begins to be
incomplete at these magnitudes and the purely visual impression may be
flawed.

%
%
%

\begin{figure*}
\begin{minipage}{150mm}
\begin{center}
\includegraphics[width=150mm]{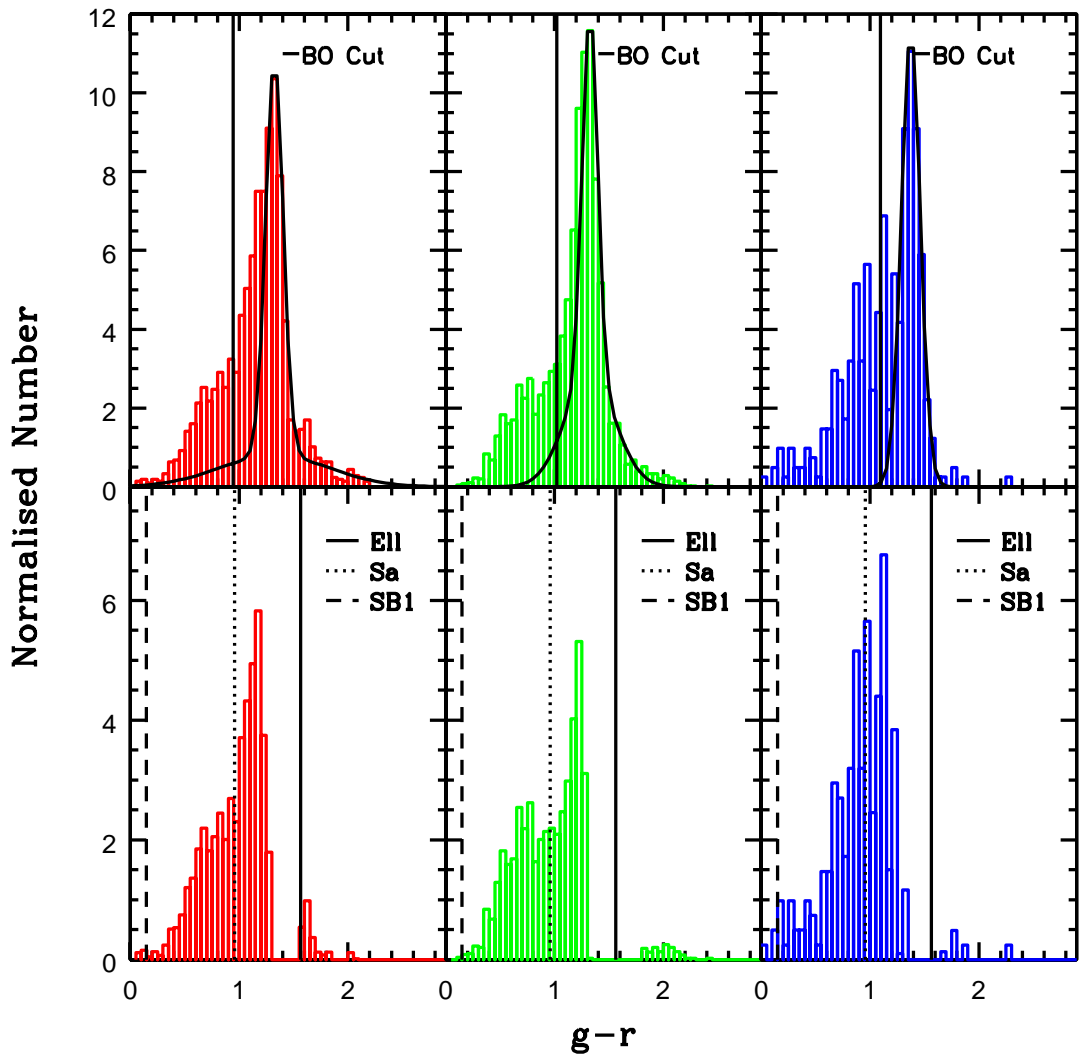}
\caption{The top panels show stacked histograms for the cool, mid, and
  hot samples before red sequence subtraction. All cluster data are
  background subtracted and $k$-corrected to $z=0.3$. The absolute
  numbers of galaxies in each bin have been re-scaled purely for
  visualisation purposes. The vertical line indicates the
  Butcher-Oemler blue cut location at $z=0.3$. The bottom panels show
  the stacked histograms after red sequence subtraction (see text for
  details). The vertical lines incidate the observed frame colours of
  the Elliptical, Sa and Starburst 1 (SB1) models of Kinney et
  al. (1996).}
\label{fig_AllPlots}
\end{center}
\end{minipage}
\end{figure*}

An alternative approach is to consider the colour distribution of all
galaxies brighter than some limit.  The colour distributions derived
from the unweighted sum of all clusters in each temperature sample
(corrected to $z=0.3$) are shown in the upper three panels of Figure
\ref{fig_AllPlots}. The limiting magnitude is $r=21.5$, approximately
equal to $M_V=-20$ at this redshift.  In order to investigate the
distribution of blue galaxies in each sample the red sequence was
modelled and removed in each case.  Following \cite{loh08} the red
wing of the red sequence was fitted using a double Gaussian model with
the mean of each Gaussian set to the location of the red sequence and
the best fitting full-width at half-maximum (FWHM) and normalisation
of each profile determined using a minimum $\chi^2$ algorithm.  The
resulting model is then reflected about the location of the red
sequence and subtracted from the corresponding colour distribution.

The mean $g-r$ colour of all galaxies bluer than the location of red
sequence for each sample was calculated and found to be 0.70, 0.675 and
0.675 for the cool, mid and hot samples respectively.  We hesitate to
place too much emphasis on the trend to observe redder mean colours in
the blue cloud for lower temperature X-ray systems.  This is
principally due to potential limitations such as a) the obvious
subtraction artefact in the mid sample red sequence subtracted
distribution and b) the broad nature of such peaks and the low number
of galxies involved (in the cool systems).

\begin{figure}
\includegraphics[width=84mm]{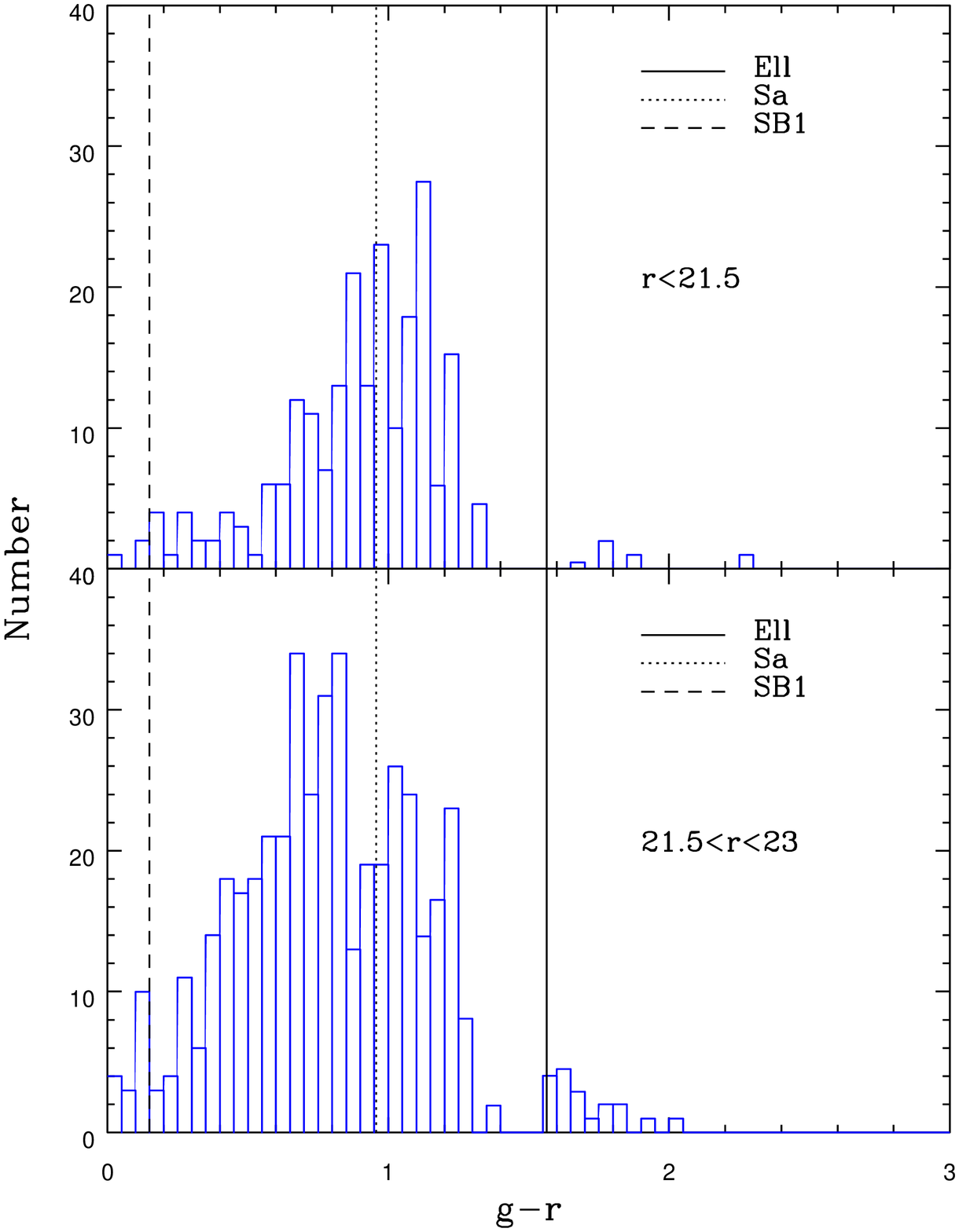}
\caption{Red sequence subtracted histogram for the cool sample. Each
  histogram is computed at $z=0.3$. The upper panel displays
  background correctd sources with $r<21$. The lower panel displays
  background corrected sources with $21.5<r<23$. The vertical lines
  incidate the observed frame colours of the Elliptical, Sa and
  Starburst 1 (SB1) models of Kinney et al. (1996).}
\label{fig_RCut}
\end{figure}

The red sequence subtracted distribution for the cool sample does
appear to show a population of extremely blue galaxies not seen in the
corresponding mid and hot cluster samples.  To quantify this
additional population we define the extremely blue fraction to be the
fraction of galaxies bluer than the colour of an Sc galaxy
($g-r=0.257$ at $z=0.3$) for all galaxies down to $r=21.5$ for each
sample.  These extreme blue fractions were found to $0.028\pm0.004$,
$0.0006\pm0.0001$ and $0.§001\pm0.0002$ for the cool, mid and hot
samples respectively.  It was noted in Figure \ref{fig_magvar} that
the fraction of blue galaxies in the cool cluster sample increases
with the increasing faint magnitude limit used to define the blue
fraction.  With this in mind we computed the extremely blue fraction
for the cool sample using galaxies in the magnitude range $21.5<r<23$
as shown in Figure \ref{fig_RCut}.  The computed extreme blue fraction
is equal to $0.052\pm0.004$, i.e. the population of extremely blue
galaxies increases with increasing magnitude.

\section{Discussion}
\label{sec_conc}

We have presented an analysis of the fraction of blue galaxies in a
large sample of X-ray clusters spanning a wide range of X-ray
temperature with uniform optical photometry.  We have noted that the
computed blue fraction values display a trend to decrease with both
decreasing redshift and increasing temperature.  

The trend of decreasing cluster blue fraction versus redshift has been
observed previously and has been interpreted as a global effect driven
by the decreasing rate of field galaxy infall onto the cluster
environment (e.g. \citealt{ellingson01}).  Coupled with the increasing
cluster mass with decreasing redshift, the specific infall rate per
unit cluster mass decreases at an even faster rate.  Infall effects
are further compounded with the sharply declining global star
formation rate as a function of decreasing redshift \citep{hopkins06}
which may be viewed as a decrease in the available gas supply of
infalling galaxies.  Looking beyond the redshift dependent component
of the Butcher-Oemler effect, the decreasing blue fraction with
increasing X-ray temperature identified in this paper is consistent
with an environmental component to the Butcher-Oemler effect.

The trend of decreasing blue fraction versus temperature is most
apparent whan comparing the across XMM-LSS and CCCP samples rather
than within either sample. It remains possible that this difference is
due in part to some subtle selection effect arising during the
construction of the two samples rather than as a result of some
physical process linked to the global environment.  However, it is not
clear what form such a bias would take.  At this point we recall that
the cool clusters may display a mild bias to higher gas densities or
more centrally concentrated systems compared to the average
population. One would naively expect such systems to be more effective
at processing infalling field galaxies compared to the average
(i.e. lower X-ray luminosity) cluster over the same temperature
interval \---\ effectively introducing a mild bias toward lower blue
fractions in this temperature sub-sample.  One could therefore argue
that access to a more complete census of the low-temperature cluster
population would only amplify the trend of blue fraction versus
temperature.  It has also been noted that optically selected clusters
display different optical properties compared to X-ray selected
samples (see Haines et al. 2009 for a discussion).  However, X-ray
samples of bright clusters are typically complete in terms of the
optical properties they sample (mainly because bright X-ray clusters
are rare and thus it is possible to select complete samples).
Moreover, clusters selected by single band optical observations
(i.e. purely based upon the projected overdensity of bright galaxies)
are more likely to be biased toward high blue fraction values due to
the high mass-to-light ratios of blue star forming galaxies.  We
therefore note that any possible bias between the XMM-LSS and CCCP
samples would be naively be expected to generate the opposite trend to
that observed in this paper and we remain confident that the trend of
blue fraction versus temperature we have identified is not a result of
sample selection effects.

A currently favoured explanation for the observed trend of blue
fraction versus temperature is that infalling field galaxies are
processed physically as they interact with the cluster environment.
This interaction may take the form of ram pressure stripping whose
effectiveness is a relatively simple function of the mass scale
represented by the group or cluster into which the galaxies are
falling.  An alternative explanation is that infalling galaxies are
processed via galaxy-galaxy interactions and therefore respond more
readily to a combination of local rather than global velocity
dispersion and galaxy density.  Whatever the cause, it is clear that
the process by which blue galaxies are processed to appear red is more
complete in hotter (more massive) environments compared to cooler
(less massive) environments.

Disentangling the above two effects is difficult as, to first order,
the effects upon star formation in the infalling galaxy depend on
environment in the same manner: ICM stripping and subsequent SF
suppression are expected to be weakest (though not absent) in group
environments \---\ galaxy populations in these environments will
display a larger fraction of blue galaxies relative to more massive
clusters \---\ while SF enhancements associated with galaxy-galaxy
interactions will be greater on group scales.  Each effect would
naively generate the same observed trend of blue fraction versus
temperature.

The blue fraction in each temperature sub-sample displays a very
similar trend versus scaled radius and is consistent with an infalling
field population albeit the cool sample is offset to higher blue
fraction \citep{ellingson01}.  However, the cool sample displays a marked
increase in blue fraction with decreasing magnitude.  The increase in
the cool sample blue fraction with increasing magnitude results from
both a moderate increase in the number of blue galaxies but also in a
relative deficit in the numbers of red galaxies compared to those in
the mid and hot samples.  We will address the red sequence luminosity
function of the cool, mid and hot samples in a subsequent
paper. However, we note that the relative deficit of red galaxies in
cool clusters compared to the mid and hot samples is consistent with
the scenario where the processing of blue galaxies to red galaxies
during infall is relatively incomplete.

Having subtracted the red sequence contribution from each colour
histogram the blue galaxy colour distribution for each temperature
sub-sample appears very similar.  We do note an excess of extremely
blue galaxies in the cool sample compared to the mid and hot
samples. The excess is small (an extremely blue fraction of 3\%
compared to $<1\%$) yet significant. In addition the fraction of
extremely blue galaxies increases in the cool sample from $\sim3\%$ to
$\sim 5\%$ for galaxies displaying $r<21.5$ and $21.5<r<23$
respectively.  One may tentatively associate these galaxies as
actively star-bursting galaxies potentially driven by galaxy-galaxy
interactions (e.g. \citealt{prop03}).  \cite{prop03} comment that the
absence of a strong Butcher-Oemler effect in near-infrared selected
cluster galaxy populations compared to optically selected populations
may be due to an increasing contribution from faint blue dwarfs whose
optical brightness is boosted by recent star formation.  The extremely
blue population identified in the cool sample is nominally consistent
with this explanation yet clearly forms only a small component of the
overall blue cluster galaxy population.  The small relative
contribution of such potentially star bursting galaxies to the cluster
population may be understood in terms of the short timescale over
which a recent star burst will affect the integrated colour of an
established stellar population.  This issue has been addressed by
\cite{barger96} who simulate the colour evolution with time of an
underlying passive (elliptical-type) and continuous star formation
(spiral-type) stellar population experiencing a short ($\sim
100$~Myr), secondary ($\sim 10\%$ by mass) burst of star formation
associated with an interaction resulting from the cluster environment.
The shift towards bluer colours as a result of a moderate burst of
star formation lasts typically only as long as the burst duration
before the luminosity weighted colour reverts to that of the
underlying pre-burst stellar population.  The effect is more
pronounced when one applies the condition that all star formation
ceases after the secondary burst, in which case the continuously star
forming stellar population (spiral-type) reddens to approximately the
same colour as the passive population within $\sim 1$~Gyr of the
burst.

Our analysis has identified a clear environmental dependence in the
blue fraction of galaxies. However, the cause of this environmental
Butcher-Oemler effect \---\ whether it be ram pressure stripping or
galaxy-galaxy interactions \---\ cannot be discerned unambiguously on
the basis of the colour distribution of the blue galaxy population in
each sample.  Clearly further information is required, most notably
the combination of colour information with morphological information.
One would naively expect the above two environmental processes to
affect galaxy morphology in markedly different ways, e.g. ram pressure
effects hould not disrupt galaxy disks whereas a strong disruptive
effect is expected from galaxy-galaxy interactions. We will present
this investigation in a future paper.

\section*{Acknowledgments}

The authors wish to thank Graham Smith, Michael Balogh, Stefano
Andreon, Pierre-Alain Duc, Florian Pacaud and Kevin Pimblett for
providing useful comments during the development of this paper. SAU
and JPW acknoweldge financial support from the Canadian National
Science and Engineering Research Council (NSERC).

\bsp

\label{lastpage}

\end{document}